\documentclass[aps,prl,twocolumn,showpacs,amsmath,amssymb,nofootinbib]{revtex4-1}
\usepackage{amsmath}
 \usepackage{color} 
\usepackage{graphicx}
\usepackage{dcolumn}
\usepackage{bm}
\usepackage{subfigure}
\usepackage{multirow}
\usepackage{makecell}
\usepackage{ltablex,booktabs}
\usepackage[bookmarksnumbered, pdfstartview=FitH,colorlinks,urlcolor=blue, citecolor=blue,linkcolor=blue] {hyperref}
\usepackage[mathlines]{lineno}
\usepackage{array}
\usepackage{amsmath}
\usepackage{booktabs}
\usepackage{soul}  
\usepackage{lineno} 
\newcommand{\BESIIIorcid}[1]{\href{https://orcid.org/#1}{\hspace*{0.1em}\raisebox{-0.45ex}{\includegraphics[width=1em]{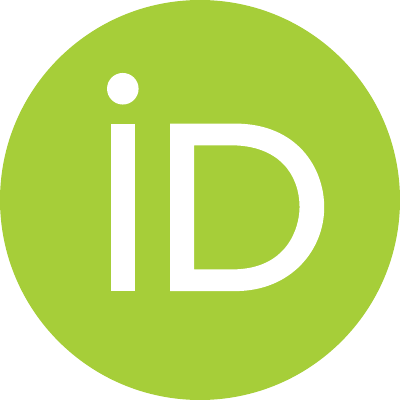}}}}

\newcommand*{\JP}{J/\psi}
\newcommand{\SigPee}{\Sigma^{+}\to pe^{+}e^{-}}
\newcommand{\SigTagone}{\bar{\Sigma}^{-}\to \bar{p}\pi^{0}}

\newcommand{\sigmap}{\Sigma^{+}}
\newcommand{\sigman}{\bar{\Sigma}^{-}}
\newcommand{\deltap}{\Delta(1232)^{+}}
\newcommand{\deltan}{\bar{\Delta}(1232)^{-}}
\newcommand{\lampi}{\bar{\Lambda}\pi^{-}}
\newcommand{\pee}{pe^{+}e^{-}}
\newcommand*{\proone}{\ensuremath{J/\psi \to \Sigma^{+}\bar{\Sigma}^{-}}}
\newcommand*{\protwo}{\ensuremath{J/\psi \to \bar{\Lambda}\Sigma^{+}\pi^{-}}}
\newcommand{\BF}{\mathcal{B}}
\newcommand{\BSigPee}{\mathcal{B}(\Sigma^{+}\to pe^{+}e^{-})}
\newcommand{\nbf}{({2.07^{+0.74}_{-0.61}}_{\rm stat.} \pm 0.10_{\rm syst.})\times 10^{-5}}

\newcommand{\njp}{(10087\pm 44)\times10^6}
\newcommand{\sif}{6.3\sigma}
\begin{document}
\preprint{APS/123-QED}

\title{Observation of Rare Semi-leptonic Decay $\SigPee$}

 \author{
 \begin{small}
   \begin{center}
M.~Ablikim$^{1}$\BESIIIorcid{0000-0002-3935-619X},
M.~N.~Achasov$^{4,c}$\BESIIIorcid{0000-0002-9400-8622},
P.~Adlarson$^{88}$\BESIIIorcid{0000-0001-6280-3851},
X.~C.~Ai$^{94}$\BESIIIorcid{0000-0003-3856-2415},
C.~S.~Akondi$^{33A,33B}$\BESIIIorcid{0000-0001-6303-5217},
R.~Aliberti$^{42}$\BESIIIorcid{0000-0003-3500-4012},
A.~Amoroso$^{87A,87C}$\BESIIIorcid{0000-0002-3095-8610},
L.~P.~An$^{54,h}$\BESIIIorcid{0000-0002-3274-5627},
Q.~An$^{83,70,\dagger}$,
M.~S.~Anderson$^{42}$\BESIIIorcid{0009-0008-1550-2632}
% ...
\\
\vspace{0.2cm}
(BESIII Collaboration)\\
\vspace{0.2cm} {\it
$^{1}$ Institute of High Energy Physics, Beijing 100049, People's Republic of China\\
% ...
$^{4}$ Budker Institute of Nuclear Physics SB RAS (BINP), Novosibirsk 630090, Russia\\
% ...
$^{33}$ INFN Sezione di Ferrara, (A)INFN Sezione di Ferrara, I-44122, Ferrara, Italy; (B)University of Ferrara, I-44122, Ferrara, Italy\\
% ...
$^{42}$ Johannes Gutenberg University of Mainz, Johann-Joachim-Becher-Weg 45, D-55099 Mainz, Germany\\
% ...
$^{54}$ Peking University, Beijing 100871, People's Republic of China\\
% ...
$^{70}$ State Key Laboratory of Particle Detection and Electronics, Beijing 100049, Hefei 230026, People's Republic of China\\
% ...
$^{83}$ University of Science and Technology of China, Hefei 230026, People's Republic of China\\
% ...
$^{87}$ University of Turin and INFN, (A)University of Turin, I-10125, Turin, Italy; (B)University of Eastern Piedmont, I-15121, Alessandria, Italy; (C)INFN, I-10125, Turin, Italy\\
$^{88}$ Uppsala University, Box 516, SE-75120 Uppsala, Sweden\\
% ...
$^{94}$ Zhengzhou University, Zhengzhou 450001, People's Republic of China\\

\vspace{0.2cm}
$^{\dagger}$ Deceased\\
% ...
$^{c}$ Also at the Novosibirsk State University, Novosibirsk, 630090, Russia\\
% ...
$^{h}$ Also at State Key Laboratory of Nuclear Physics and Technology, Peking University, Beijing 100871, People's Republic of China\\
% ...

}

 \end{center}
 \end{small}
 }

\begin{abstract}
Using $(10087\pm 44)\times 10^6$ $J/\psi$ events collected with the BESIII detector, the rare semi-leptonic decay $\Sigma^{+}\to pe^{+}e^{-}$ is studied at an electron-positron collider experiment for the first time. 
The absolute branching fraction is measured to be $({2.07^{+0.74}_{-0.61}}_{\rm stat.} \pm 0.10_{\rm syst.})\times 10^{-5}$ with a statistical significance of $6.3\sigma$.
As is typical for flavor-changing neutral-current hyperon decays, the $\Sigma^{+}\to pe^{+}e^{-}$ decay is dominated by long-distance contributions.  
The measurement of its branching fraction, combined with the branching fractions of the already observed decays $\Sigma^+\to p\gamma$
and $\Sigma^+ \to p \mu^+\mu^-$, is essential for constraining Standard Model calculations, and provides critical experimental information for the study of new physics beyond the Standard Model. 
\end{abstract}
\maketitle

\begin{figure}[htbp]
    \centering
    \begin{minipage}{0.45\textwidth}
        \includegraphics[width=\linewidth]{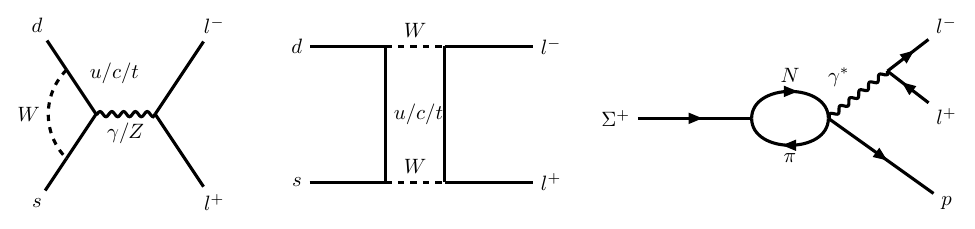}
    \end{minipage}
	\caption{Short-distance contributions from the penguin (left) and box (middle) diagrams, and an example of a long-distance contribution (right) for $\sigmap\to p l^+l^-$.
	}\label{fyman}
\end{figure}

The rare semi-leptonic hyperon decay $\sigmap\to pl^{+}l^{-}$ with $l=e, \mu$ is a flavor-changing neutral-current (FCNC) process, which receives both short-distance and long-distance contributions in the Standard Model (SM). The short-distance contributions, which arise from box diagrams, Z-penguins, and electromagnetic penguins, as shown in Fig.~\ref{fyman}, are suppressed due to cancellations inside of loops, in accordance with the Glashow-Iliopoulos-Maiani mechanism~\cite{GIM}. This leads to a combined branching fraction (BF) from all short-distance contributions that is substantially smaller than theoretical expectations for the long-distance contributions~\cite{theory2005, theory2018}. In the presence of new physics (NP) beyond the SM, the short-distance contributions could be significantly enhanced, and some of the observables in this decay might be exceptionally sensitive to signs of NP~\cite{theory2018,theory2022,theory2025}.

The dominant long-distance contribution, shown in Fig.~\ref{fyman}, relies on four complex form factors $a$, $b$, $c$, and $d$~\cite{theory2005,theory2018,theory2022,theory2025,theory1972,theory1988}. Their imaginary parts have been calculated using amplitudes for the weak non-leptonic decay $\sigmap\to(N\pi)^+$ followed by the conversion $(N\pi)^+\to p\gamma^*$, where $N$ denotes either a proton or a neutron, and $\gamma^*$ is a virtual photon. The real parts of $a$ and $b$ are extracted from the most precise measurement of the BF and decay asymmetry parameter of $\Sigma^+\to p\gamma$ from BESIII~\cite{sigmapr}, critically reducing the theoretical uncertainties in the calculation of the long-distance contribution. However, information from $\Sigma^+\to p\gamma$ alone leaves an eight-fold ambiguity in the predictions for $\Sigma^+\to pl^+l^-$, due to the inability to extract the individual signs and magnitudes of $a$ and $b$. In 2025, the LHCb Collaboration performed a high-precision measurement of the BF of the di-muon mode, reporting $\BF(\Sigma^+\to p\mu^+\mu^-)=(1.09\pm0.17)\times10^{-8}$~\cite{pmumu2025}. This result agrees well with one of the eight possible predictions within the SM~\cite{theory2025}, and the corresponding di-muon invariant-mass spectrum is also consistent with the expectation. Nonetheless, further experimental information is required to constrain the real parts of $c$ and $d$, which are currently estimated using vector-meson-dominance models~\cite{vector_meson1,vector_meson2}, and whose contributions are comparable to those from $a$ and $b$ in the di-muon channel. Thus, the predictions still suffer from ambiguity.  

The decay mode $\SigPee$ is dominated by a nearly on-shell radiative weak transition followed by Dalitz conversion~\cite{dalitz2,dalitz3}.  Compared to the di-muon mode, the di-lepton invariant mass range in the di-electron mode has a much smaller lower bound, which greatly enhances the impact of the $a$ and $b$ terms and suppresses the contributions from $c$ and $d$.  This makes the long-distance contributions particularly pronounced, resulting in a significantly less ambiguous prediction in the di-electron channel compared to the di-muon channel.   Using input from the measurements of $\Sigma^+\to p\gamma$ and $\Sigma^+\to p\mu^+\mu^-$, the BF of $\SigPee$, $\BSigPee$, is predicted to be in the range $[7.1, 7.7]\times 10^{-6}$~\cite{theory2025}. Therefore, the $\SigPee$ decay provides a unique opportunity to cross-validate the $\sigmap\to p\gamma$ measurements and offers an independent way to check the SM prediction for $\sigmap\to p\mu^+\mu^-$.
Experimentally, the only previous evidence for this decay came from the Saclay hydrogen bubble chamber experiment in 1969~\cite{pee1969}, with three candidate events leading to $\BSigPee = (7.7\pm4.6)\times 10^{-6}$~\cite{theory2018}. It is therefore timely to perform a direct measurement, as combining the results with the data on the weak radiative decay and the di-muon mode could test the consistency among various SM predictions. It is also interesting that these three modes can serve as complementary probes of CP violation within the SM and beyond~\cite{theory2026}.

In this Letter, the measurement of $\BSigPee$ is performed using $\njp$ $J/\psi$ events~\cite{jpsidata} collected at BESIII. Two $J/\psi$ decay channels, $\proone$~(Chan-I) and $\protwo$~(Chan-II), are used to produce the $\Sigma^+$ and are combined in the BF measurement.  The double-tag (DT) approach is implemented to extract signal yields and suppress background contamination. 
A data sample, referred to as the single-tag (ST) sample, is reconstructed using only the $\SigTagone$ decay in Chan-I and using only the $\lampi$ system in Chan-II.
The intermediate particles 
$\bar{\Lambda}$ and $\pi^0$ are reconstructed via the decays $\bar{\Lambda}\to \bar{p}\pi^+$ and 
$\pi^0\to\gamma\gamma$, respectively. Events in which the signal decay $\SigPee$
is reconstructed in the system recoiling against the candidates of the ST sample are denoted as DT candidates. Throughout this Letter, charge conjugation is always implied unless noted otherwise.

The design and performance of the BESIII detector are described in detail in Refs.~\cite{BESIIIdetector,BESIIIwhite}. Simulated samples are produced with a {\sc geant4}-based~\cite{Geant4} Monte Carlo (MC) package, which includes the geometric description of the BESIII detector.
For Chan-I, a MC sample of the process $\JP\to\sigmap(\to \rm anything)\sigman(\to\bar{p}\pi^{0})$ is generated according to its full helicity decay amplitude~\cite{helicity,sigmaheli}, and is used to determine the detection efficiency of the ST side. The two dominant background sources in Chan-I are $\sigmap\to p\pi^0$ and $\sigmap\to p\gamma$, and the corresponding exclusive MC samples are also generated 
according to their respective helicity decay amplitudes.  A signal MC sample with the $\Sigma^+$ decaying into the signal final state $p e^+ e^-$ is used to determine the DT detection efficiency.  The signal process is generated using the same helicity decay amplitude as $\sigmap(\to p\gamma)\sigman(\to\bar{p}\pi^{0})$ and further accounts for the four form factors arising from both long-distance and short-distance contributions, as described in Ref.~\cite{theory2018}. In this study, the real parts of $a$ and $b$, $\mathrm{Re}(a) = -12.15$ MeV and $\mathrm{Re}(b) = 4.78$ MeV~\cite{theory2025}, are selected as the nominal values.
For Chan-II, a similar strategy is used for the MC sample production, except for the three-body decay $\JP\to\sigmap\bar{\Lambda}\pi^{-}$ which is modeled using a uniform phase space distribution rather than helicity amplitudes. 
An inclusive MC sample of generic $\JP$ decays, corresponding to the luminosity of the full data set, is generated to estimate potential backgrounds.  All known decay modes are modeled with {\sc evtgen}~\cite{EvtGen1,EvtGen2} using BFs taken from the Particle Data Group (PDG)~\cite{pdg}, and the remaining unknown decays are modeled with {\sc lundcharm}~\cite{Lund}. Final-state radiation from charged final-state particles is incorporated using {\sc photos}~\cite{Photons}.

Charged tracks detected in the helium-based main drift chamber (MDC) are required to be within a polar angle range of $|\!\cos\theta| < 0.93$, where $\theta$ is defined with respect to the beam direction.
For charged tracks originating directly from the $J/\psi$ decay, the distances of the closest approach to the interaction point (IP) are required to satisfy $|V_z| <$10~cm along the beam direction and $|V_r| <$1~cm in the plane perpendicular to the beam. For tracks from $\sigman$ decay, the requirements of $|V_z| <$10~cm and $|V_r| <$1~cm are applied, while for those from $\bar{\Lambda}$ decay, only $|V_z| <$20~cm is required. Particle identification (PID) is implemented by combining measurements of the specific ionization energy loss in the MDC ($\mathrm{d}E/\mathrm{d}x$) and the flight time in the time-of-flight system (TOF), and each charged track is assigned a particle type of pion or proton according to the highest probability. 

Photon candidates are identified using showers in the electromagnetic calorimeter (EMC). The deposited energy of each shower must be more than 25~MeV in the barrel region ($|\!\cos\theta| \le 0.80$) or more than 50~MeV in the end-cap region ($0.86 \le |\!\cos\theta| \le 0.92$). To suppress electronic noise and showers unrelated to the event, the difference between the EMC time and the event start time is required to be within [0, 700]~ns. Each $\pi^0$ candidate is reconstructed using a pair of photons with invariant-mass in the interval [0.116, 0.148]~GeV/$c^2$. To improve the resolution, a kinematic fit is performed by constraining the invariant mass of the photon pair to the $\pi^0$ mass~\cite{pdg} and requiring the corresponding $\chi^2$ of the fit to be less than 200. The four-momenta updated by the kinematic fit are used in the subsequent analysis. 
The $\bar{\Lambda}$ candidates are reconstructed via the decay $\bar{\Lambda} \to \bar{p}\pi^+$ and a vertex fit~\cite{vertex} is performed on each $\bar{\Lambda}$ candidate. In events with multiple $\bar{p}\pi^+$ combinations, the candidate with the least $\chi^2$ from the fit is kept.

\begin{figure}[htbp]
    \centering
    \begin{minipage}{0.235\textwidth}
        \includegraphics[width=\linewidth]{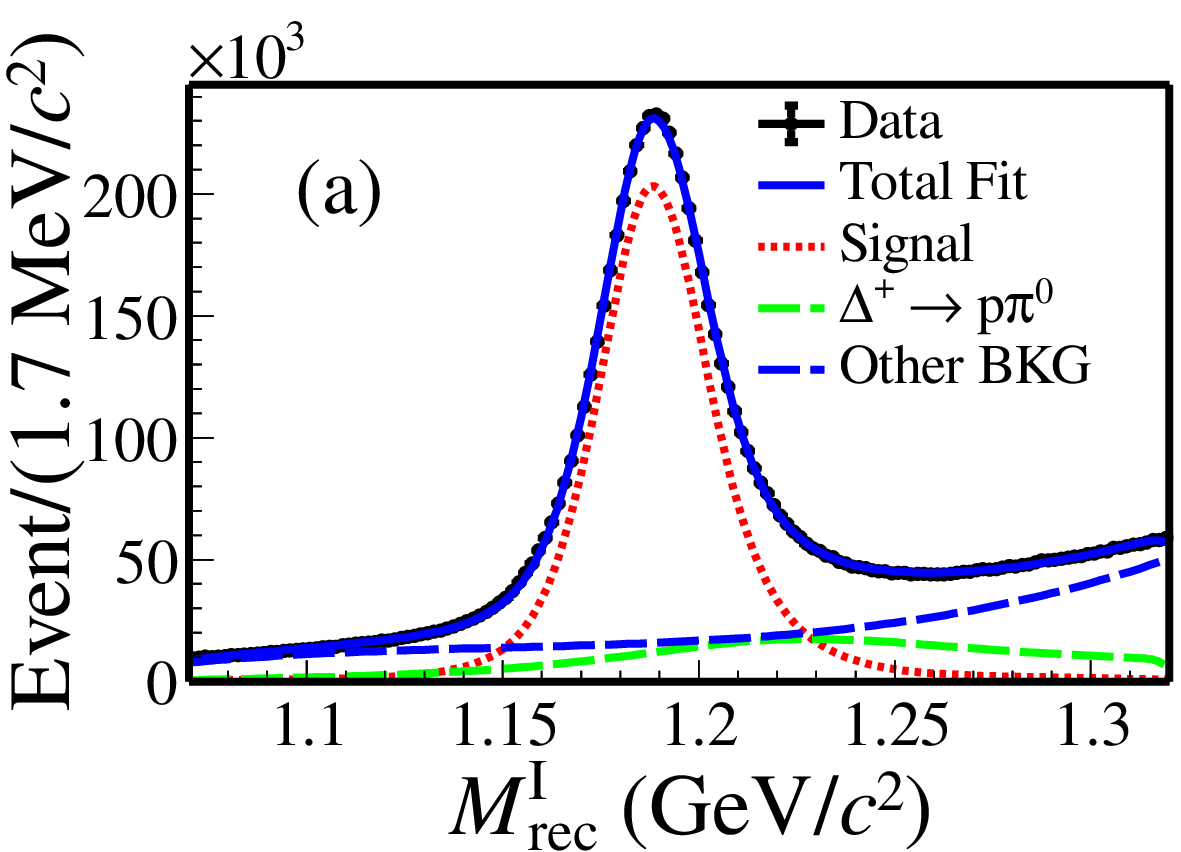}
    \end{minipage}
    \hfill
    \begin{minipage}{0.235\textwidth}
        \includegraphics[width=\linewidth]{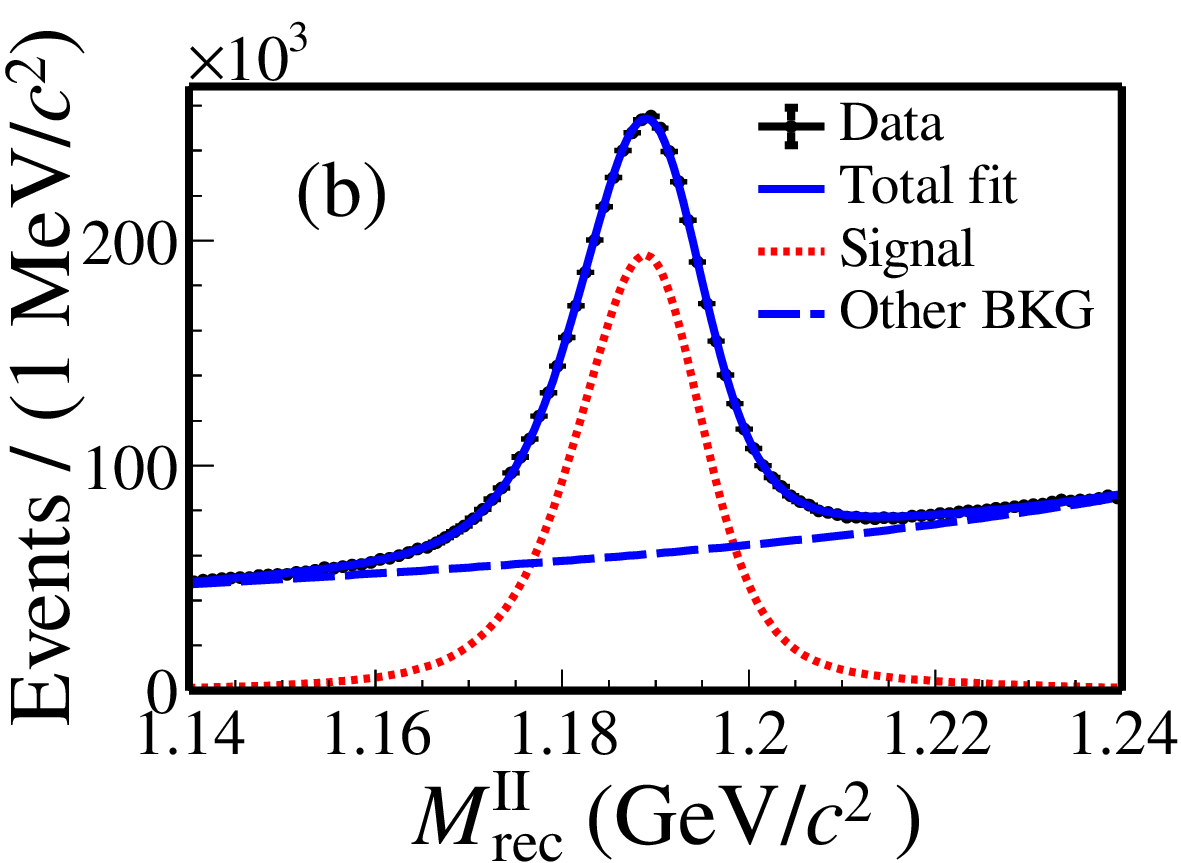}
    \end{minipage}
	\small\caption{Fits to the $M^{\rm I}_{\rm rec}$ and $M^{\rm II}_{\rm rec}$ distributions of ST candidates for (a) Chan-I and (b) Chan-II. The black points with error bars are data and the blue solid curves are the total fit results. The red dotted, blue dashed, and green dashed curves show the shapes of the signal, other non-peaking backgrounds, and the $\JP\to\deltap(\to \rm anything)\deltan(\to\bar{p}\pi^{0})$ background components, respectively.}\label{ST_fitting}
\end{figure}

The ST candidates in Chan-I are identified using the invariant mass of the system recoiling against the reconstructed $\SigTagone$ candidates, $M^{\rm I}_{\rm rec} \equiv \sqrt{(E_{\rm cms}-E_{\bar{p}}-E_{\pi^{0}})^2/c^4 - (\bold P_{\bar{p}}+\bold P_{\pi^{0}})^2/c^2}$, where $E_{\rm cms}$ is the center-of-mass energy, $E_{\bar{p}} (\bold P_{\bar{p}})$ and $E_{\pi^{0}} (\bold P_{\pi^{0}})$ are the energies (momenta) of the anti-proton and $\pi^{0}$ in the $\JP$ rest frame, respectively. The $\sigman$ candidates are required to satisfy the requirement 
$|M_{\bar{p}\pi^{0}}-M_{\sigman}|<13.5$~MeV/$c^2$, where $M_{\bar{p}\pi^{0}}$ is the invariant mass of the selected anti-proton and $\pi^0$, and $M_{\sigman}$ is the nominal $\sigman$ mass~\cite{pdg}. If there is more than one combination of 
$\bar{p}\pi^{0}$ satisfying the requirements above, the one with the minimum $|M_{\bar{p}\pi^{0}}-M_{\sigman}|$ is kept. 
The distribution of $M^{\rm I}_{\rm rec}$ is shown in Fig.~\ref{ST_fitting}~(a), where the dominant background $\JP\to\deltap(\to {\rm anything})\deltan(\to\bar{p}\pi^{0})$ forms a broad peak under the evident $\Sigma^+$ signal. A binned maximum likelihood fit is performed on the $M^{\rm I}_{\rm rec}$ distribution to determine the ST yield, and the fit result is presented in Table~\ref{Fitting_table}. In the fit, the signal and $\JP\to\deltap(\to {\rm anything})\deltan(\to\bar{p}\pi^{0})$ background are described by their MC simulated shapes convolved with a Gaussian function, which represents the resolution difference between data and MC simulation.
Other non-peaking backgrounds are described by a third-order polynomial. 

Similarly, the ST candidates in Chan-II are identified using the invariant mass of the system recoiling against the reconstructed 
$\lampi$ candidates, $M^{\rm II}_{\rm rec} \equiv \sqrt{(E_{\rm cms}-E_{\bar{\Lambda}}-E_{\pi^-})^2/c^4 - (\bold{P}_{\bar{\Lambda}}+\bold{P}_{\pi^-})^2/c^2}$, where $E_{\bar{\Lambda}} (\bold P_{\bar{\Lambda}})$ and $E_{\pi^-} (\bold P_{\pi^-})$ are the energies (momenta) of the $\bar{\Lambda}$ and $\pi^{-}$ in the $\JP$ rest frame, respectively. The $\bar{\Lambda}$ and $\pi^-$ tracks are fit to a common vertex\cite{vertex}. If there is more than one combination of $\lampi$, the one with the minimum $\chi^2$ from the vertex fit is kept. The updated four-momenta after the fit are used in the subsequent analysis, and the invariant mass of the anti-proton and $\pi^+$, $M_{\bar{p}\pi^+}$, is required to satisfy $1.105<M_{\bar{p}\pi^+}<1.125$~GeV/$c^2$. The distribution of $M_{\rm rec}^{\rm II}$ is shown in Fig.~\ref{ST_fitting}~(b), where there is a clear $\Sigma^+$ signal. A binned maximum likelihood fit is performed on the $M_{\rm rec}^{\rm II}$ distribution to determine the ST yield, and the results are presented in Table~\ref{Fitting_table}. In the fit, the signal shape is described by MC simulated events convolved with a Gaussian function, and the non-peaking background is modeled by a third-order polynomial.

The signal decay $\SigPee$ is searched for among the remaining tracks recoiling against the  ST candidates. At least three charged tracks are required, and the requirements on their distances to the IP are the same as those for tracks from the $\bar{\Sigma}^-$ decay. Protons and pions are identified using the PID approach, and all others are assigned as electrons or positrons based on their charge without applying additional PID criteria. A vertex fit~\cite{vertex} is applied to constrain the $\pee$ candidates to originate from a common vertex, and the updated four-momenta are used in the subsequent analysis. 

Additional requirements are imposed in Chan-I to improve the significance of the signal. If there are multiple $\pee$ combinations, only the one with minimum $\Delta E \equiv |E_{\pee}-E_{\rm cms}/2|$ is retained, where $E_{\pee}$ is the energy of the selected $\pee$ candidate in the $J/\psi$ rest frame. The $\Delta E$ of the candidates is also required to satisfy $\Delta E<10$~MeV. To suppress contamination from the decays of $\Delta(1232)$, events within the range 1.175 $< M_{\rm bc}^{\rm rec} <$ 1.205~GeV/$c^2$ are selected, where $M_{\rm bc}^{\rm rec}$ is defined as the beam-constrained mass recoiling against the $\bar{\Sigma}^-$, $M_{\rm bc}^{\rm rec} \equiv \sqrt{E_{\rm cms}^2/4c^4 - (\bold P_{\bar{p}}+\bold P_{\pi^{0}})^2/c^2}$. In addition, the $p\bar{p}$ vertex is reconstructed using a secondary vertex fit approach~\cite{vertex}, and the background due to $\Delta(1232)$ decays is further rejected by the requirement $L/\sigma_{L}>2$~\cite{LLerr}, where $L$ is the distance between the $p\bar{p}$ vertex and the IP and $\sigma_{L}$ stands for its resolution. The $L/\sigma_{L}$ requirement exploits the lifetime difference between the $\sigmap$ and $\Delta(1232)$ and reduces the number of $\Delta(1232)$ background events by 95$\%$ while retaining 66$\%$ of signal events. A requirement of $M_{p\bar{p}\pi^{0}}^{\rm rec}<0.1$~GeV/$c^2$ is applied to suppress the remaining background events of $\sigmap\to p\pi^{0}$, where $M_{p\bar{p}\pi^{0}}^{\rm rec}$ is defined as the invariant mass of the system recoiling against the selected particles $p\bar{p}\pi^{0}$. This requirement reduces the number of background events by 85$\%$ while preserving 88$\%$ of signal events. The background originating from $\gamma$-conversion, in which a photon undergoes material-induced conversion into an electron–positron pair, exhibits the same final state as the signal. A requirement of $R_{xy}<2.0$~cm is applied to effectively suppress this background, achieving a 98$\%$ rejection rate while retaining 47$\%$ of the signal. The variable $R_{xy}$ is defined as the distance projected onto the transverse plane between the primary IP and the $\gamma$-conversion vertex, which is reconstructed with the $e^+e^-$ pair following the algorithm in Ref.~\cite{gamconv}. Finally, a kinematic fit is utilized to constrain the selected final-state particles to the initial $\JP$ four-momentum under energy-momentum conservation.

Additional requirements are also applied in Chan-II. 
A requirement of 1.175 $<M^{\rm II}_{\rm rec}<$ 1.205~GeV/$c^2$ is used to suppress combinatorial background.   
A kinematic fit under energy-momentum conservation to the initial $\JP$ four-momentum is applied to all $\pee$ combinations, and the ST candidate for 
$\bar{\Lambda}\pi^{-}$. In the case of multiple candidates for the $\pee$ combinations, the candidate with the minimum $\chi_{4C}^2$ from 
the kinematic fit is retained for further analysis. 
To suppress the background $\JP\to\gamma\Lambda\bar{\Lambda}$, the invariant mass of the proton from the $\Sigma^+$ and the $\pi^-$ from the $\JP$ is required to be greater than 1.125~GeV/$c^2$.
Similar to Chan-I,  the background due to decays of the $\Delta(1232)$ particles is reduced
by the requirement on the distance between the $p\bar{p}$ vertex and the IP, $L/\sigma_{L}>2.0$,
which is implemented using the secondary vertex tool~\cite{vertex}.
Another requirement, $M_{\bar{\Lambda}\pi^{-}p}^{\rm rec}<0.1$~GeV/$c^2$, is used to reject the remaining background of 
$\sigmap\to p\pi^{0}$, where the variable $M_{\bar{\Lambda}\pi^{-}p}^{\rm rec}$ is defined as the invariant mass of the system 
recoiling against the selected particles $\bar{\Lambda}\pi^{-}p$. Backgrounds arising from $\gamma$-conversion are also highly suppressed by $R_{xy}<2$~cm.

\begin{figure}[htbp]
    \centering
	\begin{minipage}{0.23\textwidth}
        \includegraphics[width=\linewidth]{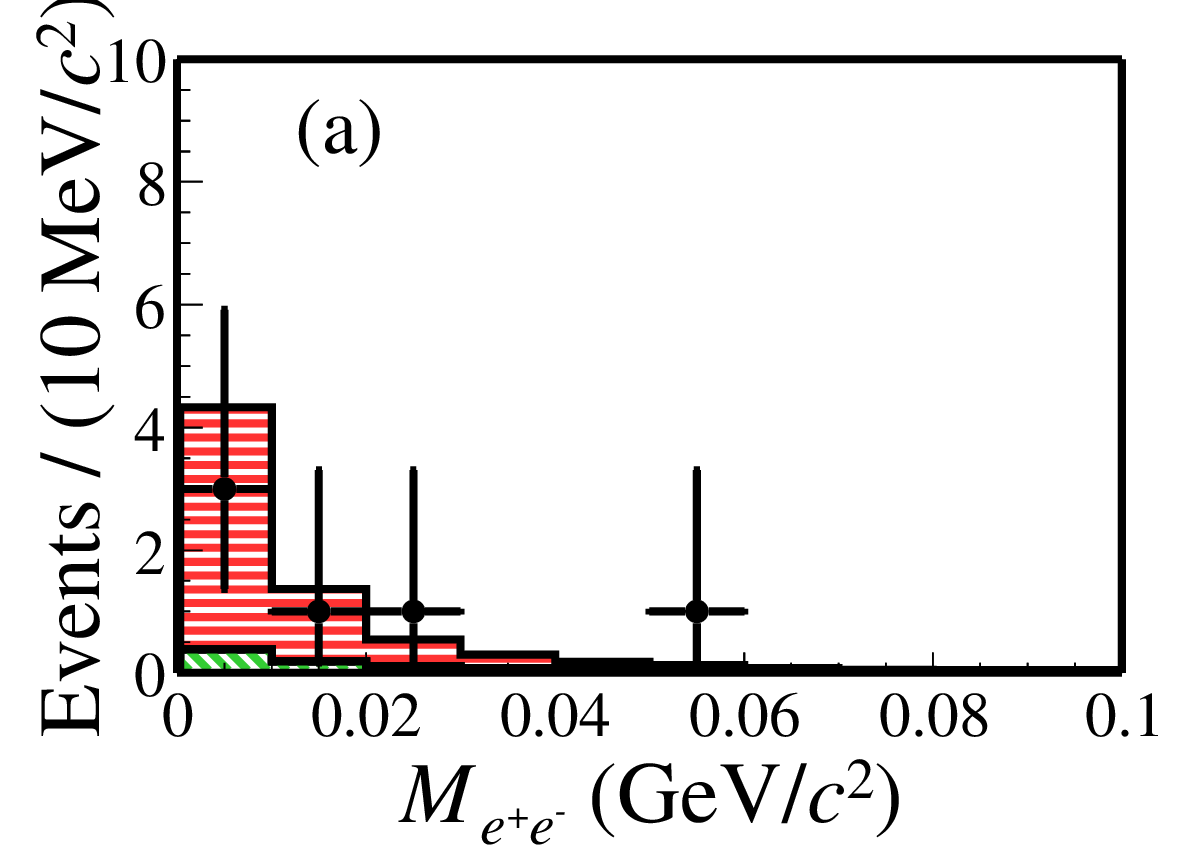}
    \end{minipage}
    \hfill
    \begin{minipage}{0.23\textwidth}
        \includegraphics[width=\linewidth]{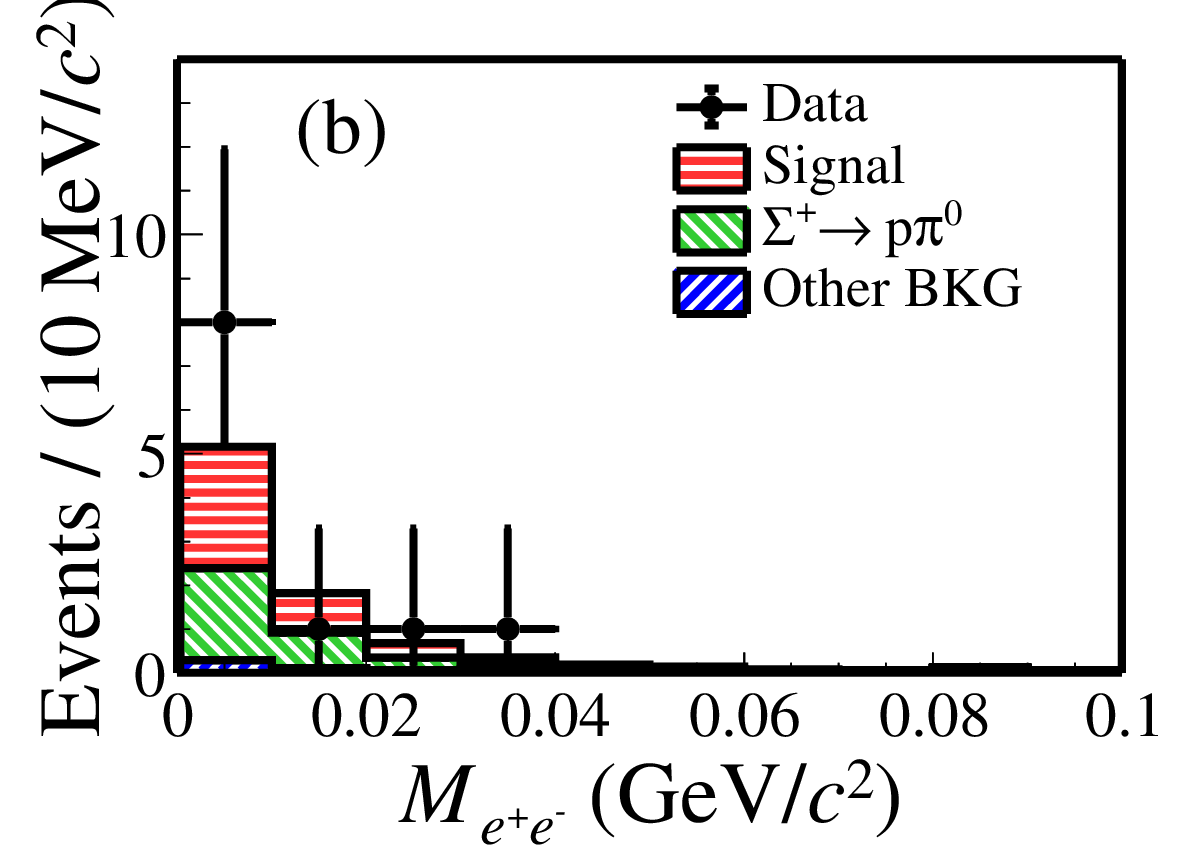}
    \end{minipage}
    \begin{minipage}{0.23\textwidth}
        \includegraphics[width=\linewidth]{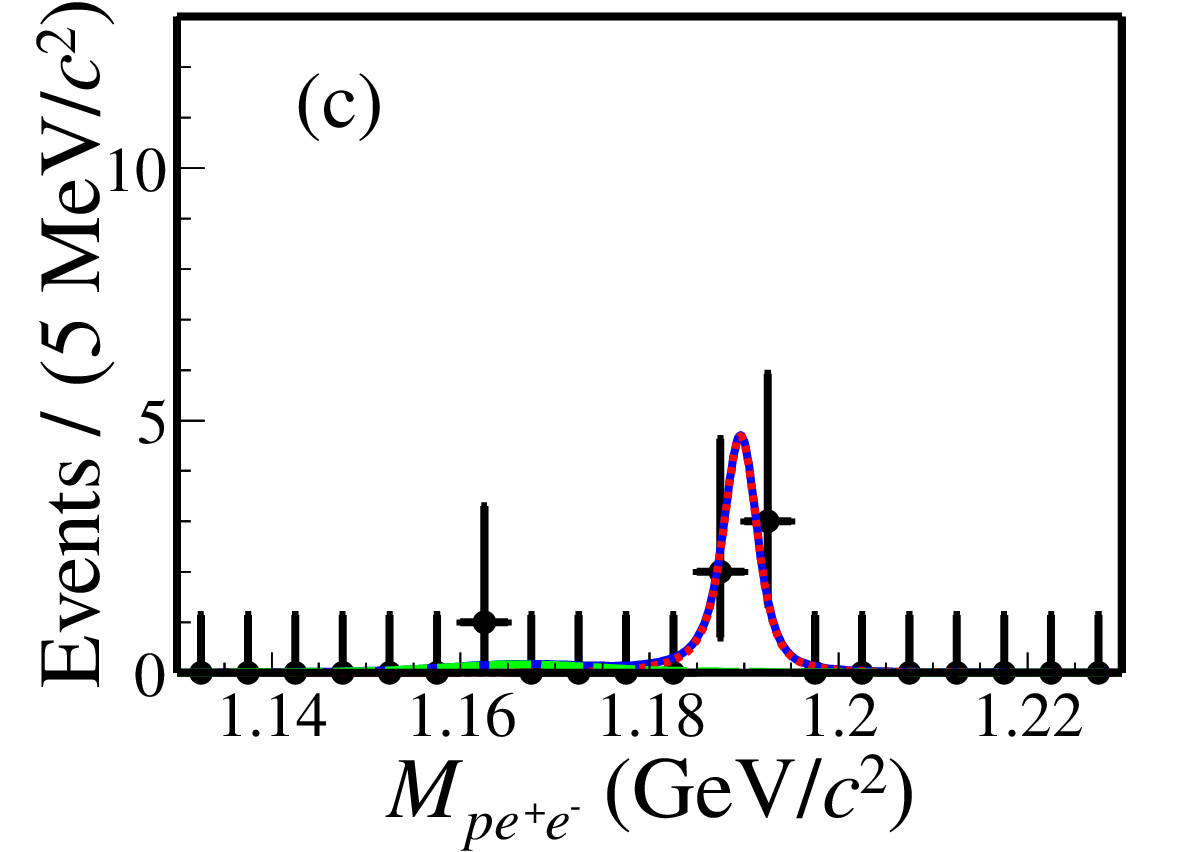}
    \end{minipage}
    \hfill
    \begin{minipage}{0.23\textwidth}
        \includegraphics[width=\linewidth]{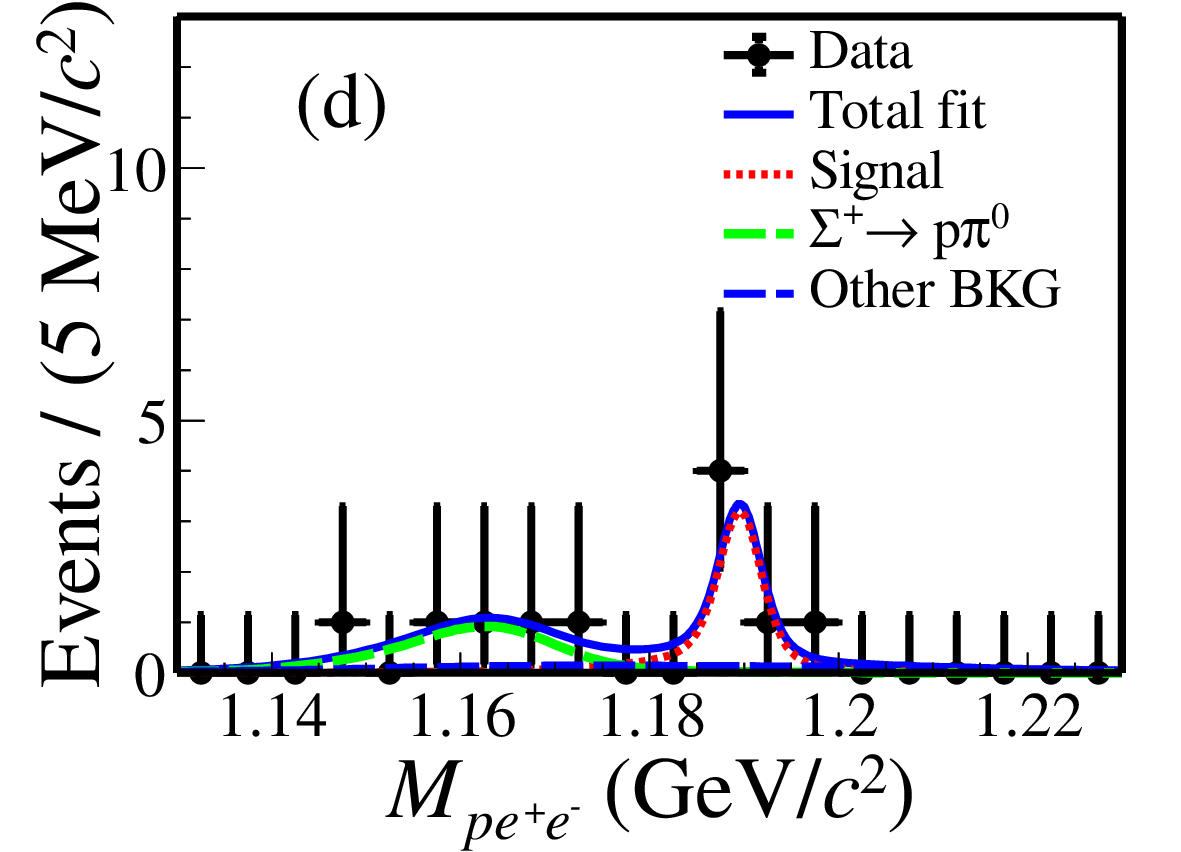}
    \end{minipage}
	\small\caption{The $M_{e^+e^-}$ distributions for (a) Chan-I and (b) Chan-II candidate events, with corresponding fits to the $M_{\pee}$ distributions shown in (c) Chan-I and (d) Chan-II. In the fit, the black points with error bars are data and the blue solid curves are the total fit results. The red dotted, green dashed and blue dashed curves show the shapes of the signal, the background $\sigmap\to p\pi^{0}$ with $\pi^{0}\to\gamma e^+e^-$, and other background components, respectively.
	}\label{DT_fitting}
\end{figure}

After applying all the above selection criteria, the resulting distributions of the invariant masses $e^+e^-$ and $\pee$, $M_{e^+e^-}$ and $M_{\pee}$, are shown in Fig.~\ref{DT_fitting}. 
The accumulation of events around $1.19$~GeV/$c^2$ in the $M_{\pee}$ distributions indicates the signal decay $\SigPee$. The background contamination has been studied with the exclusive MC samples for the two dominant background sources $\sigmap\to p\pi^{0}$ and $\sigmap\to p\gamma$. The events around $1.16$~GeV/$c^2$ originate from the background $\sigmap\to p\pi^{0}$ with $\pi^{0}\to\gamma e^{+}e^{-}$, which is consistent with the prediction from the corresponding MC samples. 
The undetected photon in the decay $\pi^{0}\to\gamma e^{+}e^{-}$ causes the invariant mass of the $p e^+e^-$ system to exhibit a smaller value compared to the nominal $\Sigma^+$ mass~\cite{pdg}. The contribution from the $\Sigma^+\to p\gamma$ decay is negligible according to the studies performed with dedicated exclusive MC and inclusive MC samples. For Chan-I, the contribution from other $J/\psi$ decay processes is negligible, while a small contribution from other $J/\psi$ decay processes is observed in Chan-II.

To obtain $\BSigPee$, a simultaneous unbinned extended maximum likelihood fit is performed on the $M_{\pee}$ distributions for Chan-I and Chan-II according to the formula 
\begin{equation*}
	N_{\rm DT} = \BSigPee(\frac{N_{\rm ST}^{\rm I}\epsilon_{\rm DT}^{\rm I}}{\epsilon_{\rm ST}^{\rm I}}+\frac{N_{\rm ST}^{\rm II}\epsilon_{\rm DT}^{\rm II}}{\epsilon_{\rm ST}^{\rm II}}),
\end{equation*}
where $N_{\rm ST(DT)}$ and $\epsilon_{\rm ST(DT)}$ denote the observed ST (DT) yields and the corresponding detection efficiencies, and the superscripts represent Chan-I and Chan-II. In the fit, $\BSigPee$ and the number of background events are unconstrained parameters. The signal shape is described by the MC simulation. The background shape of $\sigmap\to p\pi^0$ is derived from the MC simulated events, $\JP\to\sigmap(\to p\pi^{0})\sigman(\to\bar{p}\pi^{0})$ with $\pi^{0}\to\gamma e^+e^-$ for Chan-I, and $\JP\to\sigmap(\to p\pi^{0})\bar{\Lambda}(\to\bar{p}\pi^{+})\pi^-$ with $\pi^{0}\to\gamma e^+e^-$ for Chan-II. The other background shape is obtained from the inclusive MC.
The fitting results are shown in Fig.~\ref{DT_fitting} and listed in Table~\ref{Fitting_table}. 
The statistical significance of $\SigPee$ is $\sif$, which is calculated from the change of the likelihood values between fits with and without the signal component, taking into account the change in the number of degrees of freedom.

\begin{table}[htbp]
\renewcommand{\arraystretch}{1.5}
\small\caption{The observed ST yields ($N_{\rm ST}^{\rm obs}$), DT yields ($N_{\rm DT}^{\rm obs}$), ST detection efficiencies ($\epsilon_{\rm ST}$), DT detection efficiencies ($\epsilon_{\rm DT}$), and $\BSigPee$ for Chan-I and Chan-II. The first and second uncertainties are statistical and systematic, respectively.}
\label{Fitting_table}
\centering
\begin{tabular}{l r @{$\;\pm\;$} l @{\hspace{1.5em}} r @{$\;\pm\;$} l}
\hline
\hline
Mode 					& \multicolumn{2}{c}{Chan-I} 		& \multicolumn{2}{c}{Chan-II} \\
\hline
$N_{\mathrm{ST}}^{\mathrm{obs}}$ ($\times10^3$) 		& 4847.2  & 3.6    & 3729.8 & 4.2 \\
$N_{\mathrm{DT}}^{\mathrm{obs}}$  		& \multicolumn{2}{c}{$6.2^{+3.0}_{-2.4}$}    & \multicolumn{2}{c}{$4.4^{+2.4}_{-2.0}$} \\
$\epsilon_{\mathrm{ST}} (\%)$ 			& 44.55    & 0.06   & 46.18 & 0.06 \\
$\epsilon_{\mathrm{DT}} (\%)$ 			& 2.74    & 0.02   & 2.65 & 0.02 \\
\hline
$\BSigPee$ 					& \multicolumn{4}{c}{$(2.07^{+0.74}_{-0.61}  \pm 0.10)\times 10^{-5}$}\\
$\BF_{\mathrm{SM}}(\SigPee)$ 		& \multicolumn{4}{c}{(0.71 -- 0.77)$\times 10^{-5}$} \\
\hline
\hline
\end{tabular}
\end{table}

The systematic uncertainties for $\BSigPee$ comprise those associated with the ST yields, tracking and PID, the generator, and the requirements to suppress background. The DT approach on which the measurement is based ensures that uncertainties associated with the ST selection efficiency cancel. The systematic uncertainties associated with proton tracking (1$\%$) and PID (1$\%$) are evaluated using a control sample of $\JP\to p\bar{p}\pi^{+}\pi^{-}$~\cite{ptrack}. The uncertainties due to $e^\pm$ tracking are studied using the radiative Bhabha scattering process~\cite{etrack} ($1.4\%$ for Chan-I, $2.0\%$ for Chan-II). The uncertainties of ST yields are estimated by varying the background shapes ($1.4\%$ for Chan-I, $0.06\%$ for Chan-II). The uncertainty due to the $\Delta E$ requirement is estimated using a control sample of $\JP\to\sigmap(\to p\pi^{0})\sigman(\to\bar{p}\pi^{0})$ ($2.9\%$). The uncertainties of the $M_{\rm bc}^{\rm rec}$ and $M^{\rm II}_{\rm rec}$ requirements are determined by comparing efficiencies between data and MC simulation before and after applying the requirements, which are $1.4\%$ for Chan-I and $1.2\%$ for Chan-II, respectively. The uncertainty of suppressing the $\Sigma^+\to p\pi^0$ background is studied using a control sample of $\JP\to \omega p\bar{p}$ with $\omega$ decaying to $\gamma \pi^{0}$ ($2.3\%$). The uncertainty of the $R_{xy}$ requirement is estimated using a control sample of $\JP\to\pi^0\pi^+\pi^-$ with the $\pi^0$ decaying to $\gamma e^+e^-$~\cite{systgam} ($2.2\%$). The uncertainties due to the signal MC generator parameters ($\mathrm{Re}(a)$, $\mathrm{Re}(b)$) and amplitude formulas are evaluated by varying these parameters and changing the formula assumptions~\cite{theory2025} ($3.5\%$ for Chan-I, $4.2\%$ for Chan-II). For Chan-II, an additional uncertainty is assigned by re-weighting the $\sigmap$ angular distribution in MC according to data, and taking the change of the DT/ST efficiency ratio as the systematic uncertainty ($1.0\%$). The uncertainty from background shape in the simultaneous fit is evaluated by replacing the nominal background shapes with the final background shapes obtained from the inclusive MC sample, yielding a systematic uncertainty of $1.5\%$. The relative weights of the two processes are determined from the simultaneous fit. The total systematic uncertainty is calculated as
$
	\sigma_{\rm syst}=
	\sqrt{(\omega_{\rm I}\sigma_{\rm I})^2+
		(\omega_{\rm II}\sigma_{\rm II})^2+
		\sigma_{\rm simulfit}^2}
	=4.6\%
$,
where $\omega_{\rm I}$ ($\omega_{\rm II}$) and $\sigma_{\rm I}$ ($\sigma_{\rm II}$) are the weights and systematic uncertainties of Chan-I (Chan-II), respectively, and $\sigma_{\rm simulfit}$ denotes the uncertainty from the background variation in the simultaneous fit.

In summary, using $\njp$ $\JP$ events collected with the BESIII detector, the FCNC semi-leptonic decay $\SigPee$, which is dominated by long-distance contributions, is observed for the first time. The absolute BF of this decay is determined to be $\nbf$ with a statistical significance of $\sif$, which is approximately 2$\sigma$ above both the theoretical prediction and the current PDG upper limit. The comparison between the measurement and the SM prediction is presented in Table~\ref{Fitting_table}. The ratio $\frac{\BSigPee}{\BF(\sigmap\to p\gamma)}$ is determined to be $(2.08 \pm 0.68)\times 10^{-2}$, with $\BF(\sigmap\to p\gamma)$ taken from the BESIII measurement~\cite{sigmapr}, exceeding the prediction based exclusively on the Dalitz effect for $\SigPee$~\cite{theory2018}. The ratio $\frac{\BF(\sigmap\to p\mu^+\mu^-)}{\BF(\SigPee)}$ is calculated to be $(5.27 \pm 1.89)\times 10^{-4}$, with $\BF(\sigmap\to p\mu^+\mu^-)$ taken from Ref.~\cite{pmumu2025}, which lies below the SM prediction~\cite{theory1988}. With this measurement of $\SigPee$, the corresponding parameters in theoretical calculations can be further constrained. Moreover, combining with the results from $\sigmap\to p \gamma$ and $\sigmap\to p\mu^+\mu^-$, these measurements provide critical experimental input for probing potential new physics beyond the SM.	\\ 

\acknowledgements
The authors thank Professor Jusak Tandean for helpful theoretical discussion. 
The BESIII Collaboration thanks the staff of BEPCII (https://cstr.cn/31109.02.BEPC) and the IHEP computing center for their strong support. This work is supported in part by National Key R\&D Program of China under Contracts Nos. 2023YFA1606000, 2023YFA1606704, 2025YFA1613900; National Natural Science Foundation of China (NSFC) under Contracts Nos. 11635010, 11935015, 11935016, 11935018, 12025502, 12035009, 12035013, 12061131003, 12192260, 12192261, 12192262, 12192263, 12192264, 12192265, 12221005, 12225509, 12235017, 12342502, 12361141819, 12535005, 12475091; Guangzhou Navigation Project No. 2024A04J6334; the Chinese Academy of Sciences (CAS) Large-Scale Scientific Facility Program; the Strategic Priority Research Program of Chinese Academy of Sciences under Contract No. XDA0480600; CAS under Contract No. YSBR-101; 100 Talents Program of CAS; The Institute of Nuclear and Particle Physics (INPAC) and Shanghai Key Laboratory for Particle Physics and Cosmology; Agencia Nacional de Investigaci\'on y Desarrollo de Chile (ANID), Chile under Contract No. ANID CCTVal CIA250027; Istituto Nazionale di Fisica Nucleare, Italy; Knut and Alice Wallenberg Foundation under Contracts Nos. 2021.0174, 2021.0299, 2023.0315; Ministry of Development of Turkey under Contract No. DPT2006K-120470; National Research Foundation of Korea under Contract No. RS-2026-25486791; National Science and Technology fund of Mongolia; Polish National Science Centre under Contracts Nos. 2024/53/B/ST2/00975, 2025/58/E/ST2/00051; STFC (United Kingdom); Swedish Research Council under Contract No. 2019.04595; U. S. Department of Energy under Contract No. DE-FG02-05ER41374

\nocite{*}

\bibliographystyle{apsrev4-2}
\bibliography{paperbib}

\end{document}